# *In situ* X-ray diffraction studies of graphite oxidation reaction indicating different exfoliation mechanism than *ex site* studies

Karolis Vilcinskas, Fokko M. Mulder, Stephen J. Picken, and Ger J. M. Koper*

Department of Chemical Engineering, Delft University of Technology, van der Maasweg 9, 2629HZ Delft, The Netherlands

**Abstract**

We offer a brief overview of the solvent-based graphene production and summarize the current knowledge on the formation mechanism of graphite oxide that proceeds via graphite intercalation compounds. In addition, the results of our *in situ* X-ray diffraction investigation into this process are presented, discussed and contrasted to the findings by other authors, who employed the same oxidation protocol but examined the samples by *ex situ* X-ray diffraction.

Our results suggest that, contrary to the numerous reports by other authors, no strong crystalline order, unique to graphite intercalation compounds as well as graphite oxide, develops if they remain in concentrated acid. Furthermore, it also appears that, depending on the concentration, sulfuric acid molecules significantly weaken graphene-graphene interactions in graphite. Consequently, concentrated sulfuric acid may be a good solvent for graphene dispersions, if only there is sufficient energy input to separate the layers of graphene.

*Corresponding author. Tel: +31 (15) 278 8218. E-mail: G.J.M.Koper@tudelft.nl;
 (Ger Koper)

# 1. Introduction

The interest in graphene, the basic structural element of graphite, and its applications has not withered but has rather been steadily increasing ever since its isolation, for the first time in 2004 [1]. Due to its unique properties, such as exceptional inherent stiffness [2], impermeability to gases [3] and excellent electrical [4] and thermal conductivity [5], graphene has been the subject of investigation in a variety of scientific areas. Indeed, graphene and its derivatives have found applications in polymer composites [6], inorganic composites [7], sensors [8], photocatalysis [6], electrodes [9], catalysts [10], biomedicine [11] and electronic devices [12]. However, the properties and hence the performance of graphene-based materials are strongly influenced by the preparation method of graphene [13]. Preparation methods of high quality graphene are generally labour-intensive, expensive and produce only modest quantities of graphene and are therefore mainly used to provide the material for investigation of its intrinsic properties. On the other hand, due to the remarkable properties of graphene, large scale production is desired, and, thus, considerable effort has been put into the development of large scale production methods [14]. One method in particular, namely the chemical oxidation of graphite, offers the most viable route for obtaining relatively large quantities of graphene-like material. During the oxidation process of graphite, hydroxyl and epoxy groups along with a small number of ester and tertiary alcohol groups evolve on the basal plane of graphene sheets, whereas carbonyl, carboxy and 5-6 member ring lactols decorate the edges of graphene sheets [15]. Oxidation of graphite disrupts the $sp^2$-hybridized structure of graphene, and increases the interlayer spacing between sheets due to the introduction of the oxygenated groups, that confer dispersibility in a variety of solvents [16], most importantly water. In addition, the oxygenated groups on graphene oxide sheets offer different routes for chemical functionalization [17] that improve the compatibility with the polymer matrix in graphene-based-polymer composites.

Furthermore, the elimination of these groups allows one to obtain reduced graphene oxide [18] that possesses properties comparable to those of graphene as obtained by other methods. Although the chemical derivation of graphene has become the most widely employed method, the exact oxidation mechanism of the precursor – graphite – remains poorly understood [19].

In this study we review the solvent-based graphene preparation methods as these methods use the same precursor material that upon sonication in solvents yield graphene dispersions. We also briefly discuss the stability of such graphene dispersions. In addition, we outline the chemical oxidation methods and the prevailing knowledge on the formation of graphite oxide. Lastly, we present and discuss the rather surprising results of an *in situ* X-ray diffraction study of the graphite oxidation reaction, which has, in contrast to *ex situ* diffraction results, not yet been reported in literature.

## 2. Preparation of graphene and its precursor materials

*2.1. Solvent exfoliation of graphite*

Amongst the many graphene production methods, direct exfoliation of graphite in liquids has gained significant popularity [20, 21]. The formation of cavitation bubbles and the high shear forces induced by sonication of powder graphite in a solvent provide enough energy to separate the layered structure of graphite and yield mono-, bi- and few-layer defect-free graphene sheets [22]. Although the yield of such procedures is low, $\leq 0.01$ mg mL$^{-1}$ depending on the solvent [23], it can be increased by prolonging sonication time and/or increasing ultrasound power at the expense of the size of graphene sheets [22]. A recent study [24] has demonstrated that similar concentration graphene dispersions can be produced by shear exfoliation of graphite powder in N-methyl-2-pyrrolidone. Here, the change in graphite concentration, shearing time, speed, and solvent volume enable to change the concentration of the resulting graphene dispersion. Whether shearing or sonication of graphite powder in solvents, the experimental evidence has led to conclude that those solvents, such as N-methyl-2-pyrrolidone and/or *N,N*-dimethylformamide, that have a surface energy of about 70 mJ m$^{-2}$, yield the highest concentration, up 1.2 mg mL$^{-1}$ depending on the sonication time and power, graphene dispersions. Accordingly, it has been proposed that the surface energy of such solvents matches that of the graphene sheets, and as a result, is able to counter-balance the attractive van der Waals forces between the sheets, thereby preventing the exfoliated graphene sheets from aggregating [25]. However, the high boiling point and toxicity of the most studied solvents have prompted investigation into the use of milder systems. Exploration of low boiling point organic solvents, such as 1-propanol [26], chloroform [27] or acetonitrile [28], and/or solvent exchange strategies, which involve transferring graphene dispersions obtained in a high boiling point solvent to a low boiling point solvent [29], have only acquired limited success. Indeed, since the enthalpy of vaporization of a solvent is directly related to the surface energy of the solvent, only the high boiling point solvents are advantageous to prepare relatively concentrated graphene dispersions.

Another strategy to produce stable and aqueous graphene dispersions has been to employ surfactants [30-32], polymers [33, 34] or pyrene derivatives [35, 36]. The use of surfactants or polymers enables the reduction of the interfacial surface tension between graphene sheets and solvent molecules, thus promoting the stability of graphene dispersions. Ionic surfactants stabilize graphene sheets by van der Waals and/or hydrophobic interactions between the hydrophobic tails of the surfactant molecules and the graphene sheets while the hydrophilic head groups dissociate in water. Thus, graphene sheets become charged, and the associated electrical double layer repulsion ensures the stability of such dispersions [30]. Non-ionic surfactants, on the other hand, balance graphene sheets by steric effects of protruded hydrophilic tails whereas hydrophobic tails attach to the graphene sheets by van der Waals and/or hydrophobic interactions [32]. The steric effects are also thought to stabilize graphene dispersions in the presence of macromolecules. Stabilization of graphene dispersions by surfactants or polymers allows achieving higher concentrations of about 0.1-0.2 mg mL$^{-1}$, however the excess of stabilizer potentially has an adverse effect on its properties for the final graphene applications. Stabilization of graphene dispersions with pyrene derivatives has so far allowed the highest concentrations, up to 1 mg mL$^{-1}$, graphene dispersions, thus making it a very promising method. In this case, it is thought, graphene sheets are stabilized by π- π interactions between basal planes of graphene sheets and stabilizer molecules. If pyrene derivatives are decorated with electron withdrawing groups, such as sulfonic acid, donor-acceptor interactions also contribute to the stability of graphene dispersions.

Overall, due the simplicity of the method, direct exfoliation of graphite in solvents is an attractive way to obtain graphene dispersions. However, its application still remains limited by the number of suitable solvents and the resulting low graphene concentrations. Exfoliation of graphite in aqueous solutions containing surfactant or polymer molecules can increase the concentration of graphene dispersions, but the excess of stabilizer molecules can impair their properties, such as the electrical conductivity, of final graphene products.

*2.2. Graphene from other graphitic compounds*

It has long been known that lamellar compounds of graphite – so called graphite intercalation compounds (GIC) – can be produced by (electro)chemical oxidation of graphite powder in concentrated acids [37], although recently Kovtyukhova and co-authors have reported [38] that non-oxidative intercalation of some mineral acids, such as $H_3PO_4$ or $H_2SO_4$, in between graphene sheets of graphite, is possible at ambient conditions. A preceding study by Moissette et al. [39] showed that sulfuric acid-GICs could be obtained without supplying oxidizing agents, however the compound was produced at elevated temperatures where the acid decomposed yielding $SO_3$ species that are capable of oxidation. During the classical oxidation process of graphite in concentrated acids, before yielding the final product – graphite oxide – anions as well as neutral acid molecules intercalate between the layers of graphene forming GICs of various degrees of intercalation. In order to initiate the formation of these compounds, a small amount of oxidizing agent or the application of a voltage is often used to initiate the electron transfer reactions leading to intercalating species [40]. GICs are classified by their stage index, which denotes the number of graphene layers between adjacent intercalate layers [41]. Thus, in the first stage (n=1) GIC, the intercalating species fill every gallery between graphene sheets, whereas in higher stage GICs, the intercalating species occupy the interlayer space between the adjacent graphene sheets in alternate fashion leaving a certain number of unfilled galleries (see Figure 1).

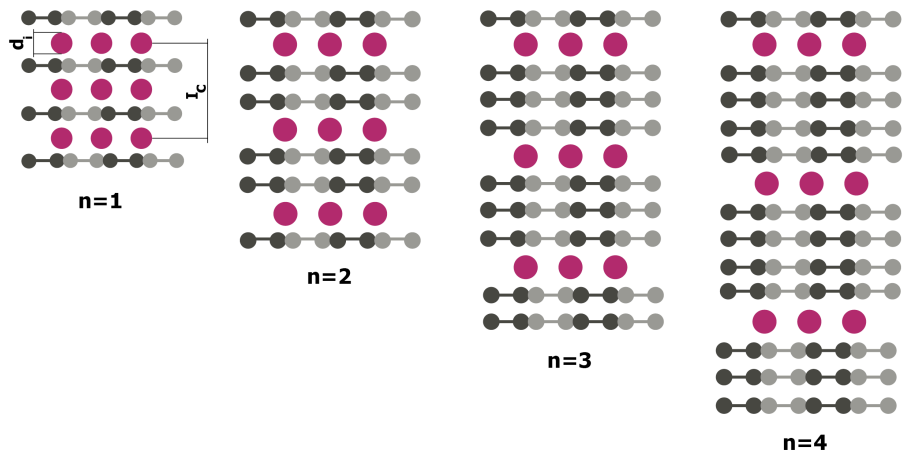

**Figure 1.** Schematic illustration of lamellar graphite intercalation compounds. Dark dots and lines represent graphene sheets in graphite whereas bigger red circles – intercalated species.

As the stacking order of graphene layers is not compromised during the intercalation process, GICs can be analysed by means of X-ray diffraction [41, 42]. By the application of Bragg's Law [43], the stage of intercalation $n$ can be obtained using the following relation [44]:

$$I_c = nc_0 + d_i \qquad (1)$$

where $I_c$ is the repeat distance, $c_0$ – the distance between the adjacent graphene sheets (3.35 Å), and $d_i$ – the size of the intercalating molecules.

In order to obtain GICs electrochemically, graphite as the anode material is immersed in a concentrated acid bath and a Platinum wire is used as the cathode. Upon application of voltage, electrons from graphite are withdrawn and flow to the counter electrode. As suggested by Metrot [45], upon removal of an electron, the potential energy of graphite rises and when it reaches a certain threshold value, anions as well as neutral acid molecules can intercalate between the layers of graphite forming GICs of different stages. The formation of GICs by electrochemical oxidation of graphite where samples were taken out of the reaction mixture at various times during the process has been extensively investigated by many authors [46-49], including direct monitoring of the intercalation process [50]. Recently, the method has also been employed in the production of graphene dispersions [51, 52].

Since the production methods [53-56] of chemically derived graphene involve graphite oxidation in concentrated acid with strong oxidants, it is natural to surmise that the oxidation step is preceded by the formation of GICs. Indeed, in the early 1940s Hofmann and Rüdorff [41] investigated GICs produced in concentrated sulfuric acid by various oxidizing agents. X-ray diffraction analysis of the samples revealed formation of GICs of various stages, generally proceeding from the higher stage to the lower, before yielding graphite oxide. In addition, they also suggested the sulfuric acid-GIC composition to be about $C_{24}^+ HSO_4^-$. Later studies by other authors [57-59] corroborated the findings, and the formula of the sulfuric acid-GIC compound has been refined to be $C_{(21-28)}^+ HSO_4^- \cdot 2.5H_2SO_4$ [44]. However, it is worth to point out that these studies were conducted *ex situ*. As the GICs are only stable in concentrated acids and readily decompose if exposed to water (vapor) [41], the analysis of these compounds is generally done by taking a small amount of the sample during the oxidation process, wrapping it in a plastic film to be analysed in an X-ray diffractometer. No studies in the literature have reported on *in situ* investigation into the structural changes of graphite during the chemical oxidation process as yet.

## 3. Experimental

### 3.1. *Sample preparation*

For the *in situ* studies, 0.1 gram of fine graphite (Gr) powder (Fluka) was dispersed in 4 millilitres of concentrated (≥95%) sulphuric acid ($H_2SO_4$, Sigma Aldrich) followed by the addition of different amounts of Potassium permanganate ($KMnO_4$, Sigma Aldrich), see Table 1 below. The samples were vigorously mixed, a small amount (approximately 0.5 mL) was quickly transferred to the specially built sample holder and measured. The same procedure was repeated for the samples dispersed in different concentrated acid solutions (constant Gr mass; 2.5 m/v % in an acid solution) or containing different amounts of Gr (constant acid concentration; 95 wt% $H_2SO_4$), see Table 2 for the detailed compositions.

Table 1. Sample compositions in in situ graphite oxidation reaction analysis.

| *Sample name* | *Mass ratio of Gr: $KMnO_4$* |
| --- | --- |
| No oxidation (S-1) | 1:0 |
| Little oxidation (S-2) | 1:1.2 |
| Partial oxidation (S-3) | 1:2.3 |
| Full oxidation (S-4) | 1:3.5 |

Table 2. Different sample compositions to evaluate acid or graphite (Gr) concentration effects.

| Sample composition[1] | Molar ratio of Gr:$H_2SO_4$ | Sample composition[2] | Molar ratio of Gr:$H_2SO_4$ |
|---|---|---|---|
| 50 wt% $H_2SO_4$ | 0.29 : 1.00 | 2.5 w/v % | 0.11 : 1.00 |
| 80 wt% $H_2SO_4$ | 0.15 : 1.00 | 6.3 w/v % | 0.28 : 1.00 |
| 90 wt% $H_2SO_4$ | 0.12 : 1.00 | 12.5 w/v % | 0.57 : 1.00 |
| 95 wt% $H_2SO_4$ | 0.11 : 1.00 | 25.0 w/v % | 1.13 : 1.00 |

[1] constant graphite concentration in acid (2.5 m/v %)

[2] constant acid concentration (95 wt%)

3.2. *Characterization*

The sample holder was specially designed and consisted of two sealable circular stainless steel plates. In the middle of the bottom plate, a circular rostrum of a few millimetres height and 1 centimetre in diameter was designed with the inlet cover, made of the high-density polyethylene to order to protect stainless steel from the chemicals. In the middle of the top plate a window, covered with polyimide film, was installed thus enabling the X-ray beam to penetrate into the sample.

X-ray diffraction (XRD) measurements in Bragg–Brentano reflection mode were performed by a PANalytical X'Pert Pro PW3040/60 diffractometer with Cu Kα radiation operating at 45 kV and 40 mA in an angular 2θ range of 5°–50° at 25°C.

**4.1. Results**

Before undertaking an analysis of the mixture samples, we have investigated X-ray diffraction patterns of pure graphite (Gr), Potassium permanganate ($KMnO_4$) and the empty sample holder, see Figure 2.

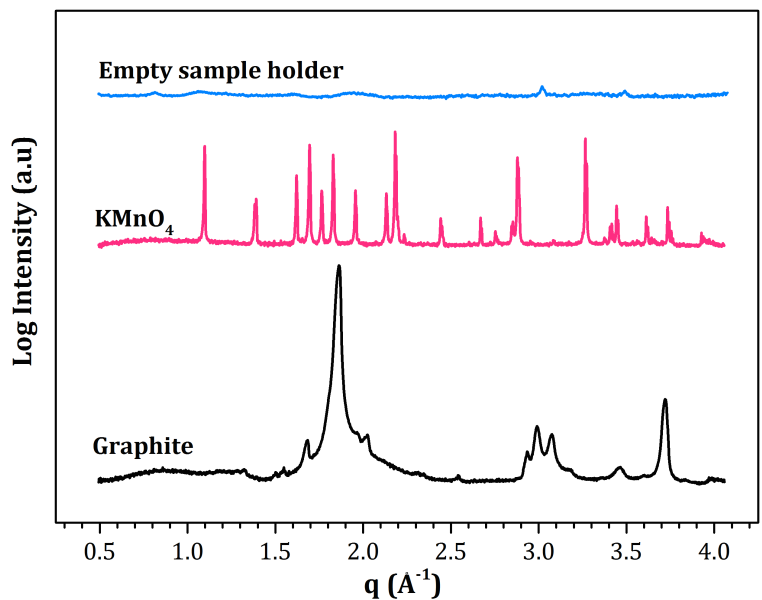

**Figure 2.** X-ray diffractograms of graphite (bottom), Potassium permanganate (middle) and empty sample holder (top). Note that the ordinate axis is in the $^{10}$log scale.

As presented in the Figure 2, the X-ray diffraction pattern of the Gr powder sample shows the characteristic peaks at 1.9 Å$^{-1}$, 3.0 Å$^{-1}$, 3.1 Å$^{-1}$ and 3.7 Å$^{-1}$ that correspond to (002), (100), (101) and (004) reflections, respectively [60]. As for the $KMnO_4$ powder sample, it exhibits numerous reflections due to its more complex arrangement of atoms [61]. The empty sample holder, on the other hand, shows no significant crystalline structure, which is characteristic for Kapton® polyimide film [62]. After we collected the X-ray diffraction patterns of the powder samples, we set out to examine the X-ray scattering of the Gr/$H_2SO_4$ mixtures containing different amounts of $KMnO_4$ by continuously measuring the samples for prolonged periods of time, see Figure 3.

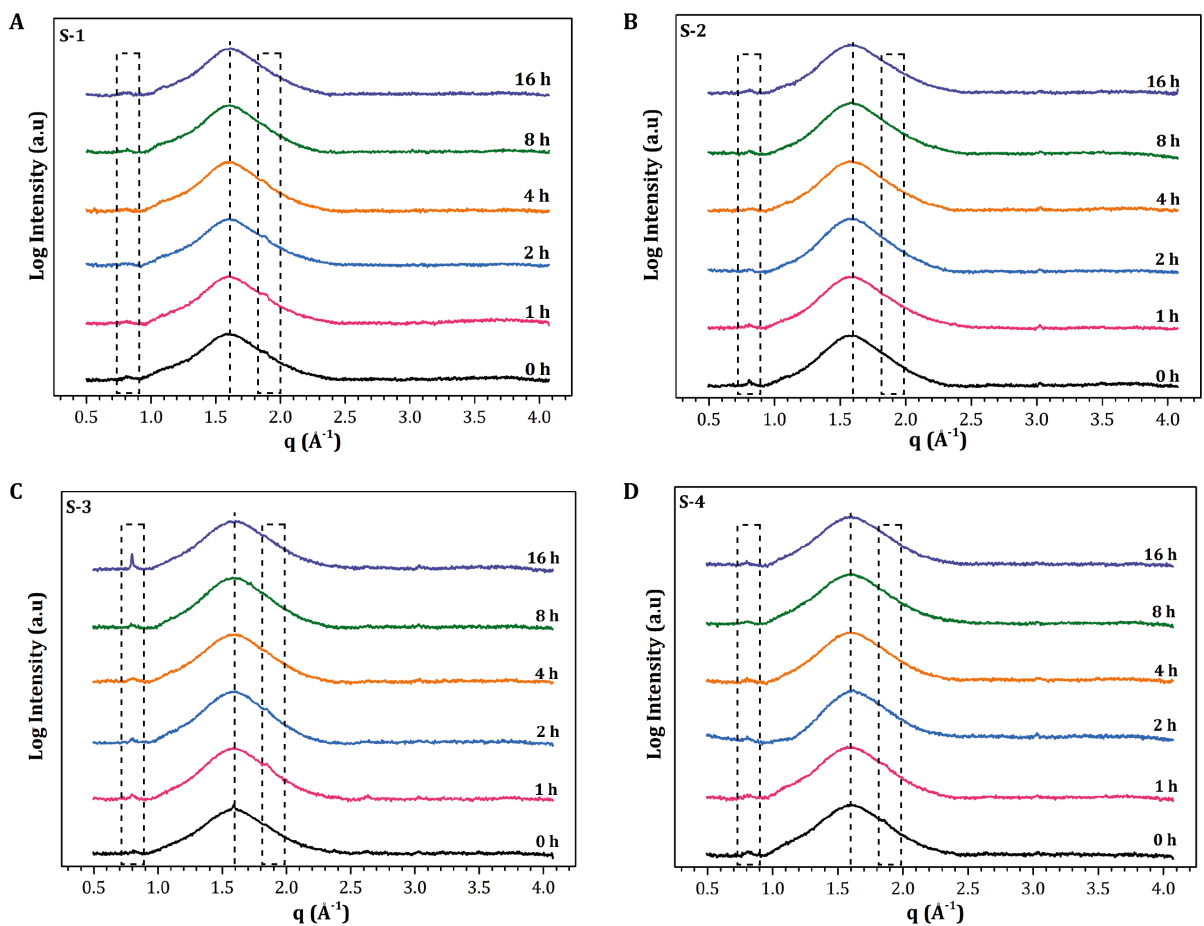

**Figure 3.** X-ray diffractograms of graphite (Gr) and sulphuric acid mixture with different amounts of added Potassium permanganate ($KMnO_4$): (A) in the absence of $KMnO_4$; (B) mass ratio Gr: $KMnO_4$=1:1.2; (C) mass ratio Gr: $KMnO_4$=1:2.3; and (D) mass ratio Gr: $KMnO_4$=1:3.5. The samples were measured at various times. The dotted lines draw attention to the most significant positions in the diffractograms.

As seen in Figure 3 A, upon addition of Gr in $H_2SO_4$, the intensity of the prominent Gr peak at 1.9 Å$^{-1}$ significantly decreased, became broader and gradually disappeared for longer times, whereas other characteristic reflections of Gr became invisible. On the other hand, the broad and intense peak at 1.6 Å$^{-1}$ appeared and remained apparent at all times as has the weak peak at 0.8 Å$^{-1}$. Upon addition of a small amount of $KMnO_4$ (mass ratio Gr: $KMnO_4$=1:1.2) to the Gr /$H_2SO_4$ mixture (Figure 3 B), the prominent Gr peak at 1.9 Å$^{-1}$ disappeared altogether, and a strong broad peak at 1.6 A$^{-1}$ along with the weak peaks at 0.8 Å$^{-1}$ and 3.0 Å$^{-1}$ were visible at all times. With further addition of KMnO$_4$ (mass ratio Gr: KMnO$_4$=1:2.3), (Figure 3 C), the peak at 1.6 Å$^{-1}$ remained prominent, the intensity of the peak at 0.8 Å$^{-1}$ slightly increased and the peak at 3.0 Å$^{-1}$ diminished. In addition, a new weak peak at 2.6 Å$^{-1}$ arose, however it remained visible only at certain times. The characteristic Gr peak at 1.9 Å$^{-1}$, albeit broadened and weakened, was also initially visible, but disappeared later. Finally, as shown in Figure 3 D, with completed addition of KMnO$_4$ as prescribed in the Hummer's method (mass ratio Gr: KMnO$_4$=1:3.5) to the Gr/H$_2$SO$_4$ mixture, the peaks at 0.8 Å$^{-1}$ and 1.6 Å$^{-1}$ remained visible with no observable development of new peaks.

In order to gain a better understanding of the origin of the observed peaks, we analysed a series of samples, containing graphite in varying H$_2$SO$_4$ concentration (2.5 w/v % graphite in acid). In addition, we also varied the concentration of graphite in concentrated H$_2$SO$_4$.. All samples were measured within 20 minutes after preparation, and their X-ray diffractograms are presented in Figure 4.

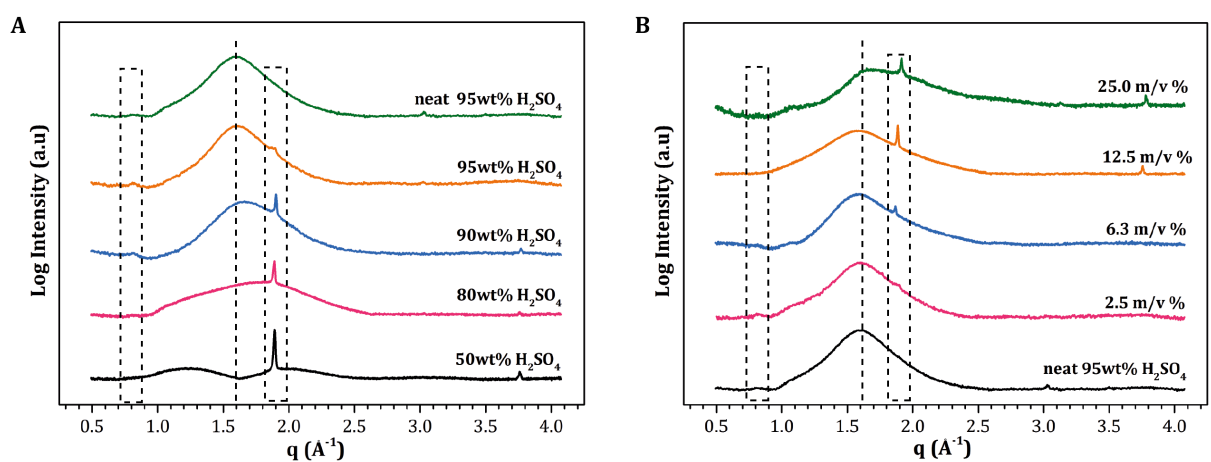

**Figure 4.** X-ray diffractograms of (A) graphite in different concentration H$_2$SO$_4$ acid solutions (2.5 w/v % graphite in acid), and (B) of various amounts of graphite in concentrated H$_2$SO$_4$. The dotted lines draw attention to the most significant positions in the diffractograms.

As seen in Figure 4 A, with increasing acid concentration, the broad peak between 1.5 Å$^{-1}$-2.0 Å$^{-1}$ began to emerge and eventually centered at 1.6 Å$^{-1}$ for high concentrations of H$_2$SO$_4$. On the other hand, with increasing acid concentration, the intensity of the characteristic Gr peak at 1.9 Å$^{-1}$ as well as the peak at 3.7 Å$^{-1}$ gradually diminished. As illustrated in Figure 4 B, with increasing amount of Gr in concentrated H$_2$SO$_4$, the characteristic peaks at 1.9 Å$^{-1}$ and 3.7 Å$^{-1}$ became more prominent. However, the broad peak at 1.6 Å$^{-1}$ did not change its shape nor position with the amount of Gr, nonetheless, at the highest investigated Gr concentration in concentrated H$_2$SO$_4$, it skewed and moved to 1.7 Å$^{-1}$. In addition, at higher concentrations of Gr, the broad and weak peak at 1.1 Å$^{-1}$ became visible.

## 5. Discussion

We anticipated that monitoring the graphite oxidation reaction *in situ* by means of X-ray diffraction, would enable us to observe the structural evolution of graphite as discussed in the Introduction. In a recent study, employing the same graphite oxidation procedure, by Dimiev and Tour [63], it was shown that graphite oxidized in three consecutive steps. Upon addition of a small amount of KMnO$_4$ (mass ratio graphite:KMnO$_4$=1:1), the first stage sulphuric acid-GIC was formed. With further addition of oxidant, its molecules diffused into the interlayer spacing between graphene sheets, reacted with acid yielding strong oxidizing species that initiate the production of pristine graphite oxide (PGO). After the addition of 4 weight equivalents of KMnO$_4$, the authors observed a very sharp and strong XRD peak at 2θ=9.7° (q=0.7 Å$^{-1}$), which led them to propose that the stacking order of the graphene sheets during the oxidation process was preserved, and only expansion along graphite's c-axis ([002] reflection) occurred. The authors argued that H$_2$SO$_4$ molecules remained intercalated in between the galleries in PGO, and that the structure was stable as long as it was not exposed to large amounts of water. Quenching the reaction mixture with H$_2$O during the oxidation procedure lead to exfoliated PGO sheets and subsequently restacked GO sheets did no longer possess long range order along the c-axis, thus giving a much weaker XRD signal. In addition, the interlayer spacing in restacked graphite oxide (CGO) was slightly reduced. Finally, based on Raman spectroscopy and optical microscopy results, the authors observed that the formation of sulfuric acid-GIC took 5 minutes only, whereas the formation of graphite oxide takes hours. The XRD results of this study are summarized in Table 3, and contrasted to our experimental evidence in the following paragraphs. It is, however, important to mention that the authors in this study prepared the samples for XRD diffraction by taking a small amount of sample from the reaction mixture at certain times into the oxidation process, centrifuging it for 30 min, discarding supernatant, and wrapping the wet powder sample in plastic.

Table 3. Characteristic peak positions observed by Dimiev and Tour [63]. Acronyms (vs) and (s) stand for very strong and strong scattering peaks, respectively.

| Sample name | Molar ratio of Gr: KMnO$_4$ | Characteristic peak positions 2θ, ° | q, Å$^{-1}$ |
|---|---|---|---|
| Stage-1 GIC | 1.0:13.2 | 22.3 (vs) | 1.6 |
|  |  | 33.7 (s) | 2.4 |
|  |  | 45.2 (s) | 3.1 |
| TF-1[1] & TF-2[1] | 1.0:13.2 1.0:6.5 | 11.4 (s) 21.6 (s) 22.3 (s) 33.7 (s) | 0.8 1.5 1.6 2.4 |
| TF-3[1] & TF-4[2] | 1.0:4.3 1.0:3.3 | 9.7 (vs) | 0.7 |
| CGO[3] | – | 11.0 (vs) | 0.8 |

[1] intermediate compounds that are produced during the oxidation reaction
[2] pristine graphite oxide
[3] conventional graphite oxide

Our X-ray diffraction patterns, presented in Figure 3 A- D however, reveal a different picture. We do not observe the evolution of a graphite structure, only a broad peak at 1.6 Å$^{-1}$ that corresponds to the distance of 3.9 Å between the adjacent sulphuric acid molecules – the so called liquid ring – and the distance is comparable to the value reported in the literature [64]. In a recent study Kovtyukhova et al. [38] reported that sulfuric acid-GIC does not produce the characteristic diffraction pattern as observed by Dimiev and Tour so long as the compound remains in liquid. Consequently, they arrived at the conclusion that the non-oxidative intercalation of sulfuric acid molecules can only occur when the hydrogen bonding network between the acid molecules is disrupted, either by heating or evaporation, since only then the acid molecules are able to initiate the inter-layer opening of graphene sheets in graphite. However, the weak and broad characteristic graphite peak ([002] reflection) at 1.9 Å$^{-1}$ observed in the Gr/H$_2$SO$_4$ mixture in our experiments suggests that sulphuric acid molecules can presumably partially intercalate in between graphene sheets in a random fashion thus disrupting the regular stacking of graphene sheets in graphite, see Figure 5 for a schematic illustration.

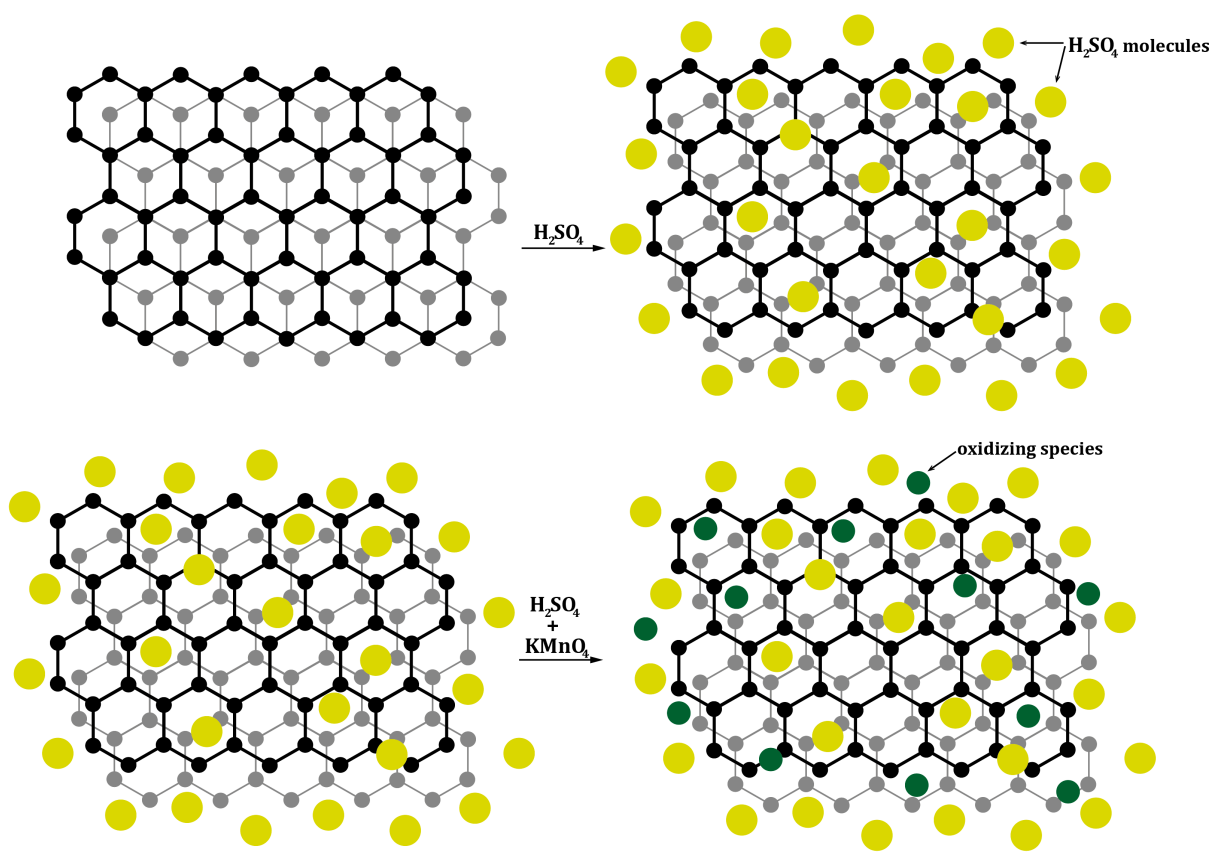

**Figure 5.** Schematic illustration of the graphite oxidation mechanism. Big dark yellow circles represent sulfuric acid molecules whereas smaller dark green circles – oxidizing species that form upon the reactions between sulfuric acid and Potassium permanganate.

This result is in agreement with the observed lowest intercalation efficiency of sulfuric acid by Kovtyukhova and co-authors [34]. When KMnO$_4$ is added, the characteristic graphite peak diminishes and/or disappears altogether It is, however, unlikely that graphite fully exfoliated into graphene sheets in the experimental procedure applied here, as this process requires high-energy input and/or time. As illustrated in Figure 4 (A), the characteristic peak of graphite 1.9 Å$^{-1}$ does not disappear for diluted H$_2$SO$_4$ solutions. This suggests that increased water content in H$_2$SO$_4$ reduces the activity of the acid molecules as they favour interactions with water molecules, and as a result the graphite-H$_2$SO$_4$ interactions are significantly reduced. On the other hand, the significant reduction of the graphite characteristic peak in concentrated sulfuric acid suggests that graphene-graphene interactions are apparently weaker compared to graphene-acid interactions. As discussed earlier, the experimental studies by Coleman and co-workers led to suggest that the solvents, which have surface energies around 70 mJ m$^{-2}$ are highly suitable for stable and relatively concentrated graphene dispersions. However, the matching surface energies may not be sufficient to guarantee a homogenous dispersion of graphene if the interfacial tension between sheets and solvent molecules is not minimized as well [65]. Now, the surface energy of sulfuric acid [66] is around the same value as that of the solvents yielding most concentrated graphene dispersions. Using a simple expression [67], the interfacial energy between two components can be estimated as follows :

$$\gamma_{AB} = \left(\sqrt{\gamma_A} - \sqrt{\gamma_B}\right)^2 \tag{2}$$

Since the surface energies of graphene and sulfuric acid are comparable, the interfacial energy between these two components is, indeed, negligible. In fact, the interfacial energy between graphite and acid is also small, assuming the surface energy value of graphite to be 52.1 mJ m$^{-2}$ [68]. This suggests, that concentrated sulfuric acid may be a suitable solvent for solvent-assisted graphene preparation as the van der Waals forces between the adjacent graphene sheets can be fully overcome.

Dried 1$^{st}$ stage sulphuric acid-GIC and PGO compounds exhibit strong and well-defined peaks, see the diffractograms in the original article [63], that are not observed in our experiments upon addition of KMnO$_4$. The absence of oxidant reflections in Figure 3 B-D indicates its reaction with concentrated H$_2$SO$_4$ that produces the oil-like, green-colored strongly oxidizing substance Mn$_2$O$_7$ [69], also observed in the reaction vial upon mixing, as well as other oxidizing species [70]. The newly developed species presumably oxidizes graphene sheets in a similar manner as proposed by Dimiev and Tour, however the 1$^{st}$ stage sulphuric acid-GIC and PGO structures are more likely to remain invisible in XRD due to the yet not defined interaction between sulfuric acid molecules and graphene sheets in graphite.

As for the weaker diffraction peak at 0.8 Å$^{-1}$ it is likely to correspond to the spacing between the second layer of sulphuric acid molecules. However, upon addition of KMnO$_4$, the intensity of the peak at 0.8 Å$^{-1}$ slightly increases, which could indicate the presence graphite oxide.

## 6. Conclusions

Our X-ray diffractometry results appear to indicate that during the graphite oxidation process using concentrated sulfuric acid and Potassium permanganate no strong crystalline order, unique to the sulfuric acid-graphite intercalation compounds and/or graphite oxide, develops if they remain in concentrated acid. This suggests that the formation of sulfuric acid-graphite intercalation compounds as well as graphite oxide cannot be excluded, but it is certain that they are not sufficiently ordered to yield the required characteristic diffraction peaks. Similarly, depending on the concentration of acid, the graphene-graphene interactions in graphite also appear to be significantly weakened by the sulfuric acid molecules, hence concentrated sulfuric acid may be a good solvent for graphene dispersions, provided there is sufficient energy input to separate the layers of graphene. However, upon removing the excess of acid, order does develop as demonstrated elsewhere.

**Acknowledgements**

This work is supported by NanoNextNL, a micro and nanotechnology consortium of the Government of The Netherlands and 130 partners. The authors thank Nico Alberts for designing and building the sample holder for X-ray diffraction measurements, and appreciate help of Kees Goubitz in collecting X-ray diffractograms. We also thank to dr.ir. M.Makkee for reviewing the manuscript.